\documentclass[conference]{IEEEtran}
\IEEEoverridecommandlockouts
\usepackage{cite}
\usepackage{amsmath,amssymb,amsfonts}
\usepackage{algorithmic}
\usepackage{textcomp}
\usepackage{xcolor}

\usepackage{graphicx}
\usepackage{subcaption}
\usepackage{orcidlink}
\usepackage{graphics}

\def\BibTeX{{\rm B\kern-.05em{\sc i\kern-.025em b}\kern-.08em
    T\kern-.1667em\lower.7ex\hbox{E}\kern-.125emX}}
\begin{document}

\title{Floor Plan-Agnostic Detection of Gait Speed Drifts Using Ambient Sensors\\
\thanks{This study is conducted as part of Innovate UK Knowledge Transfer Partnership project reference 10062173. The financial support from Legrand Care and Innovate UK is gratefully acknowledged.}
}

\author{\IEEEauthorblockN{1\textsuperscript{st} Marina Vicini}
\IEEEauthorblockA{
\textit{Aston University}\\
Birmingham, UK \\
\orcidlink{0009-0006-5622-4216}}
\and
\IEEEauthorblockN{2\textsuperscript{nd} Martin Rudorfer}
\IEEEauthorblockA{
\textit{Aston University}\\
Birmingham, UK \\
\orcidlink{0000-0001-9109-5188}}
\and
\IEEEauthorblockN{3\textsuperscript{rd} Zhuangzhuang Dai}
\IEEEauthorblockA{
\textit{Aston University}\\
Birmingham, UK \\
\orcidlink{0000-0002-6098-115X}}
\and
\IEEEauthorblockN{4\textsuperscript{th} Ahmad Beltagui}
\IEEEauthorblockA{
\textit{Aston University}\\
Birmingham, UK \\
\orcidlink{0000-0003-2687-0971}}
\and
\IEEEauthorblockN{5\textsuperscript{th} Luis J. Manso}
\IEEEauthorblockA{
\textit{Aston University}\\
Birmingham, UK \\
\orcidlink{0000-0003-2616-1120}}
}

\maketitle

\begin{abstract}
Gait speed is a vital health indicator for older adults, as changes in gait speed can reflect physiological and functional decline. Ambient sensors offer a promising, privacy-preserving solution for continuous in-home monitoring of gait speed; although it is often limited by methods requiring a home floor plan, which is frequently unfeasible. This paper proposes a novel, floor plan-agnostic method to detect gait speed drifts using only sparse ambient sensors. Our approach identifies informative sensor-to-sensor transitions and analyses fluctuations in their duration. For each sequence a non-parametric statistical test detects changes between a recent period and an initial baseline; and daily test results are aggregated to provide a robust drift detection response. We evaluate our method on a simulated dataset across four different home layouts, showing performance comparable to, and in some cases exceeding, a state-of-the-art baseline that requires floor plan information. This work demonstrates a feasible approach for scalable, cost effective gait drift detection monitoring, providing a foundation for future validation in complex real-world environments.
\end{abstract}



\begin{figure}[t]
    \centering
  \begin{subfigure}{0.15\textwidth}
    \centering
    \includegraphics[trim = 50 35 20 20, clip, width=1\textwidth]{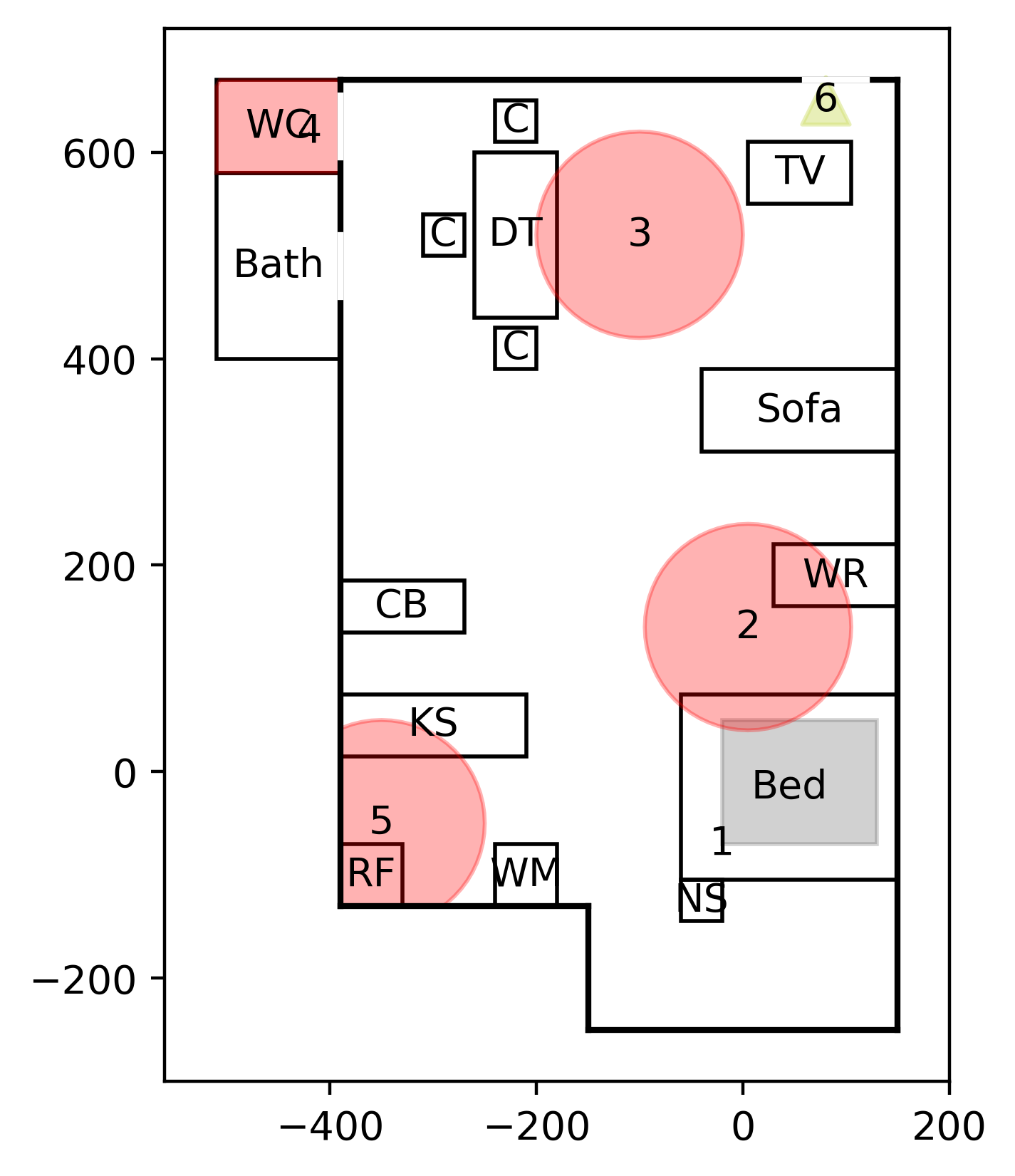}
      \caption{Layout A}
      \label{fig:layout_x_delay}
  \end{subfigure}
  \,                              
  \begin{subfigure}{0.30\textwidth}
    \centering
    \includegraphics[trim = 53 37 15 20, clip, width=1\textwidth]{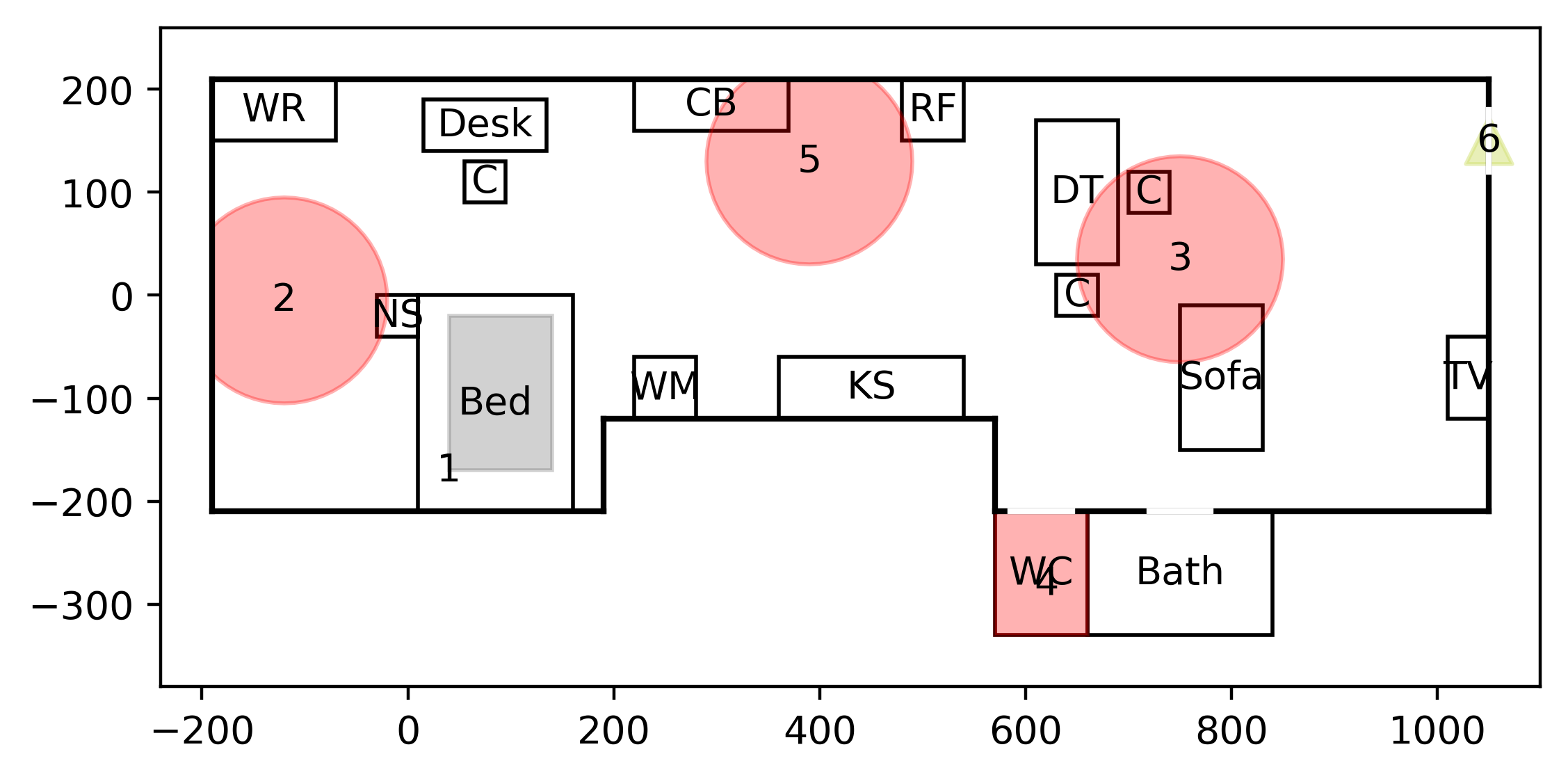}
      \caption{Layout B}
  \end{subfigure}
  \bigskip
  \begin{subfigure}{0.2\textwidth}
    \centering
    \includegraphics[trim = 50 40 20 17, clip, width=0.8\textwidth]{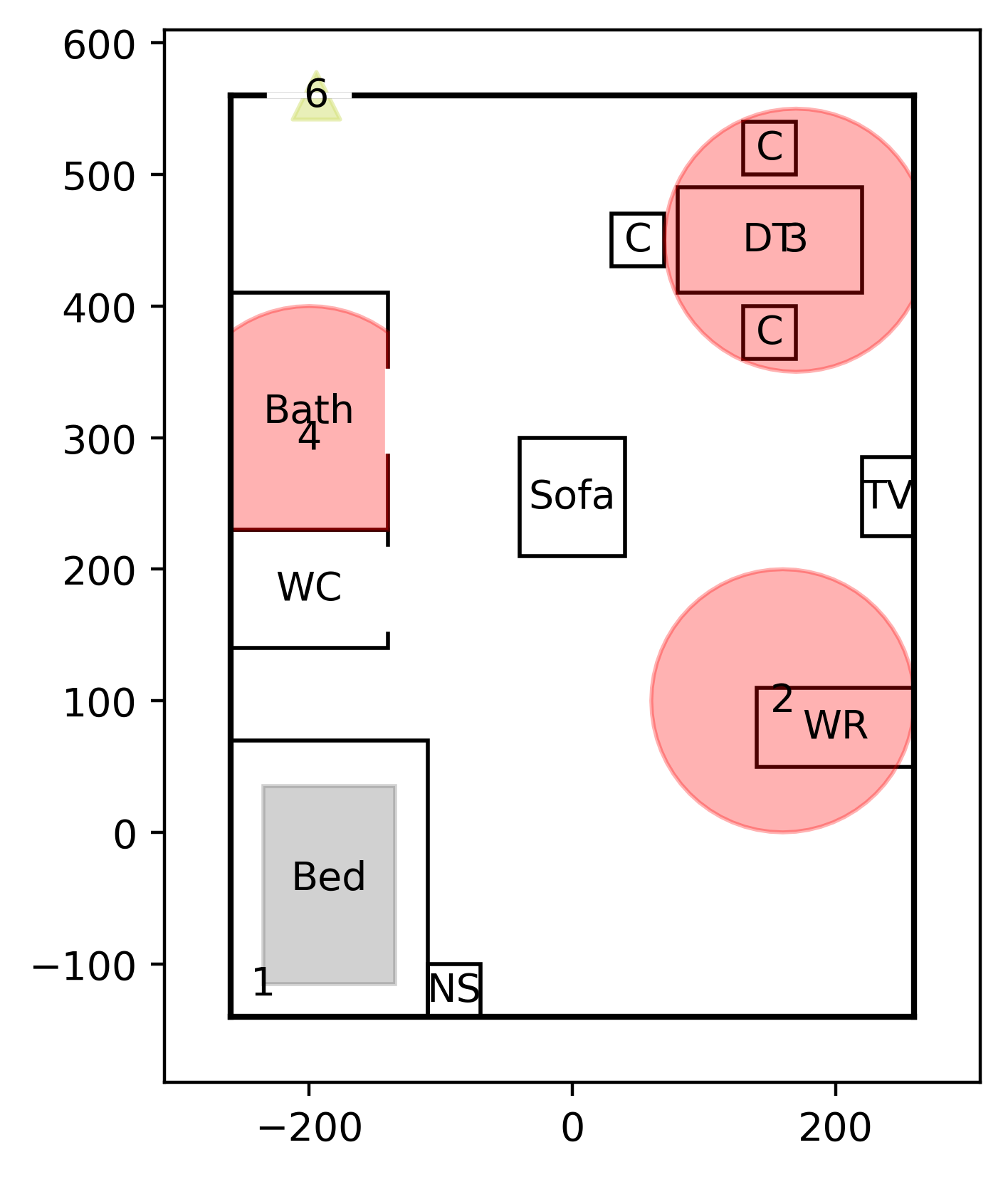}
      \caption{Layout C}
  \end{subfigure}
  \qquad
  \begin{subfigure}{0.23\textwidth}
    \centering
    \includegraphics[trim = 60 40 20 20, clip, width=1\textwidth]{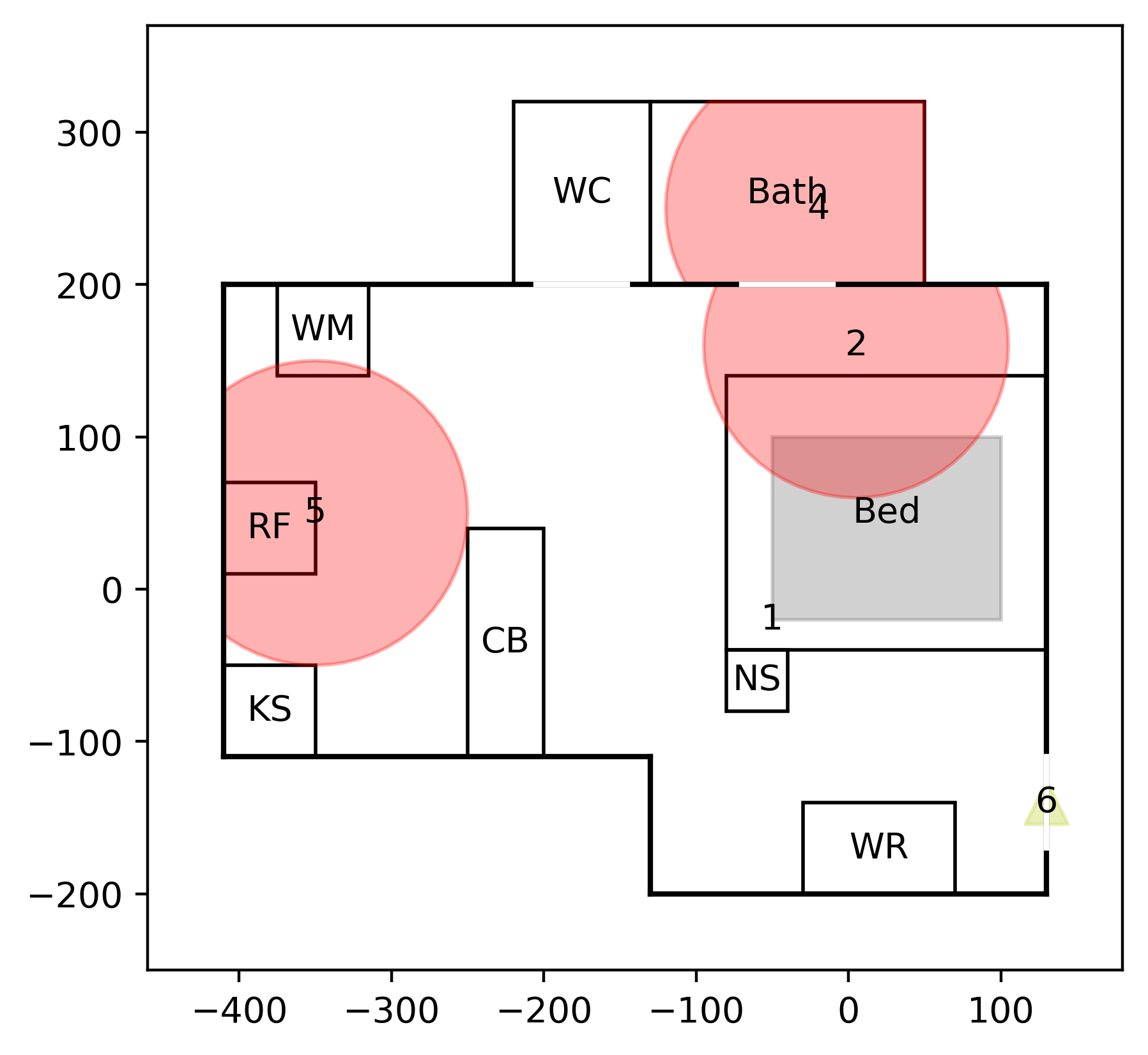}
      \caption{Layout D}
  \end{subfigure}
  \caption{The four simulated studio apartment layouts used in the experiments, generated using the simulator from~\cite{tanaka_sensor_2022, tanaka_sensor_2024}. The layouts feature different furniture arrangements and sparse sensor placements representative of commercial installations.}
    \label{fig:floorplan}
\end{figure}

\section{Introduction}
For the ageing population, mobility is a key performance indicator of health.
Gradual physical decline places older adults at risk of losing their independence and potentially needing to relocate~\cite{scheibl_when_2019}, which can compromise their right to self-determination.
Notably, regularly monitoring the health condition of older adults is crucial for detecting early-stage functional decline, as this decline shows a considerable individual variability~\cite{world_health_organization_world_2015}; it allows proactive and preventative care and making \textit{ageing-in-place} a reality for longer.


Gait is a complex activity that integrates the nervous, musculoskeletal, and cardiovascular systems~\cite{wildes_make_2019}, and a reliable measure for assessing the functional status of older adults~\cite{middleton_walking_2015}. A decline in gait speed is a consistent risk factor for adverse outcomes such as hospitalisation, mortality, cognitive impairment, and   disability~\cite{abellan_van_kan_gait_2009}, and can reflect subtle yet significant functional decline before more visible symptoms manifest. 
A change in gait speed of $0.10 \; m/s$ is widely considered a minimal clinically important difference~\cite{chui_meaningful_2012}.
Traditionally, gait speed is assessed through tests in clinical settings, such as standardised timed walks (\textit{e.g.}, a 4-metre walk)~\cite{guralnik_short_1994}. 

While valuable, these clinical assessments must be performed under the supervision of a professional, such as a clinician or physiotherapist. A significant shortcoming is their infrequent administration, providing only a snapshot of a person's health. Long-term monitoring can capture variations and gradual changes, and therefore, it has the potential for early diagnosis of various diseases and fall risk assessment, allowing for more timely interventions~\cite{montero-odasso_gait_2005}.
Home-based assessment allows for this continuous monitoring. Ambient Intelligence (AmI) and its application in healthcare, Ambient Assisted Living (AAL), leverage sensorised environments that facilitate continuous monitoring. 

A variety of sensor technologies can be used for gait analysis. Wearable sensors, such as pressure-sensing insoles~\cite{cho_walking_2023}, Inertial Measurement Units (IMUs)~\cite{ruiz-ruiz_detecting_2021}, \cite{maruyama_accuracy_2021}, or smartphones~\cite{lee_novel_2024}, can provide detailed gait metrics but require users to consistently wear and charge the devices, which is not suitable for all older adults. In contrast, ambient sensors do not require user interaction. This category includes technologies like Ultra-Wideband (UWB) or mmWave radar~\cite{lau_gait_2018, abedi_use_2019}, cameras~\cite{dolatabadi_concurrent_2016}, radio signals~\cite{hsu_extracting_2017}, and Wi-Fi sensing~\cite{wu_gaitway_2021}. However, the cost, privacy implications, or installation complexity of these systems can be barriers to widespread commercial adoption.
Motion detectors, such as Passive Infrared (PIR) sensors, and door contact sensors are less data rich but offer a cost-effective and privacy-preserving alternative for mobility assessment~\cite{sprint_unsupervised_2016}.


The motivation to using these sensors is enhanced by the proliferation of commercially available smart home kits and telecare services, such as those promoted by advisory body like the Technology Enabled Care Services Association in the UK~\cite{ADASS_unlocking_2025}, that indicate that over 2 million people in the UK use technology enabled care (TEC). These general-purpose systems are already being deployed to support older adults, typically relying on a core set of cost-effective ambient sensors like PIR motion detectors, door contacts, and smart plugs~\cite{fiorini_unsupervised_2017, aran_anomaly_2016}. Their primary objectives are broad, including abnormal behaviour detection, medication prompting, and general Activities of Daily Living (ADL) recognition~\cite{wang_survey_2023, vicini_integrating_2024, friedrich_unsupervised_2023}, rather than specialised gait analysis. Integrating gait speed drift detection capabilities into already existing systems creates a significant opportunity, as this vital health metric could be monitored at scale without requiring any new, specialised hardware.


However, the existing literature for gait speed drift detection in smart homes requires either a detailed floor plan of the home or a specific configuration of sensors in a line~\cite{hagler_unobtrusive_2010, frenken_motion_2011}. Obtaining a floor plan is not economically feasible, and creating specific setups would drive the cost for both sensors and installation. These observations lead to our central research question: Is it possible to detect gait drifts using ambient sensors without a-priori knowledge of the home's floor plan? Unlocking this capability would create a huge impact for existing and already deployed commercial systems.

This paper proposes a method to detect gait drifts in a smart home environment using only ambient motion and door sensors, without relying on a floor plan.
The contributions of this paper are as follows:
(1) a problem formulation for gait speed drift detection using sparse ambient sensor data; (2) a simulated dataset containing 4 different house layouts and a wide range of gait speed options; and (3) a method to detect gait speed drifts in a smart home without relying on any additional spatial relation between the sensors.

\section{Related Works}



To the best of our knowledge, there are no methods that aim to detect changes in gait speed from ambient sensors, without requiring any spatial context. However, there are several methods that use spatial context to directly estimate gait speed, which are most closely related to our work. These can be categorised into: (1) motion sensors placed in a line; and (2) sensors distributed across the house. 

\subsection{Sensor Array}
A common approach involves distributing an array of PIR sensors in a line, typically along a hallway ceiling or a bespoke rail, to measure the time between sensor activations. Knowing the fixed distance between sensors allows for a direct calculation of walking speed.
Hayes \textit{et al.}~\cite{hayes_unobtrusive_2009} pioneered this method by positing four motion sensors with a restricted field of view along a path of frequent traffic. Their system was validated against the GAITRite walkway system, a clinical gold standard for gait analysis, achieving a high correlation ($R^2 = 0.96$)~\cite{hagler_unobtrusive_2010, hayes_unobtrusive_2009}. However, they note that calibration is essential to account for variability in sensor response and imprecision in restricting the sensor's field of view. Similarly, Walsh \textit{et al.}~\cite{walsh_ambient_2011} used a rail-mounted system of three lined PIR sensors to derive mean daily gait velocity and intra-day circadian variations. Building on this concept, Chapron \textit{et al.}~\cite{chapron_real-time_2019} used three sensors spaced $0.6 \; m$ apart to cover a $1.20 \; m$ path. A lab-based evaluation showed a mean difference of approximately $0.09 \; m/s$ from the ground truth. To address the multi-resident challenge, their system integrated a Bluetooth-enabled wristband to associate the measured gait speed with a specific individual~\cite{chapron_real-time_2021}.

Although these lined sensor methods can be accurate, their feasibility is limited in practice due to operational or environmental reasons. The requirement for calibration and the single-purpose nature of the installation, designed exclusively for gait speed, reduce their scalability and cost-effectiveness for broader AAL applications.

\subsection{Distributed Sensors}

An alternative approach utilises sensors distributed throughout the home, which allows for multi-purpose monitoring of Activities of Daily Living (ADL)~\cite{yahaya_detecting_2021} in addition to detecting gait speed changes. These methods infer gait speed by analysing the temporal patterns of transitions between different sensor locations.

Frenken \textit{et al.}~\cite{frenken_motion_2011} developed a system that generates possible motion patterns from a 2D floor plan and sensor definitions, parsing the floor plan to identify adjacent sensors, and computing the distance of the walked path, rather than the distance of the sensors. A laboratory study found a deviation of $0.17 \; m/s$ from the baseline; the authors noted that the method is better suited for long-term trend analysis than for precise, instantaneous measurements.


Early research by Pavel \textit{et al.}~\cite{pavel_unobtrusive_2006} explored the possibility of assessing mobility without a priori geometric information. They proposed a Hidden Semi-Markov Model (HSMM) to learn the probabilistic structure of the home from sensor data, utilising the 25th percentile of transition times between inferred ``hidden states" as a proxy for speed. However, this method relies on training computationally intensive stochastic models to learn the home's topology and was limited to a preliminary evaluation on a single subject. 

Building on this, Nait Aicha \textit{et al.}~\cite{nait_aicha_continuous_2015, nait_aicha_continuous_2018} developed a method to extract valid walking paths (trajectories), identifying long and straight paths across two rooms. They observed that the durations of these paths follow non-trivial, multimodal distributions due to walking being interwoven with other activities. By fitting a mixture model (\textit{e.g.}, Poisson and Normal distributions) to these durations, they extracted the mean duration for each path to compute the corresponding gait speed. The primary advantage of this approach is its ability to be retrofitted to existing smart home installations without requiring a specific sensor layout, making it highly flexible. However, these probabilistic and topology-based methods often depend on either an explicit floor plan or a sufficiently dense and stable sensor network to learn reliable transition patterns. In contrast, we propose a method that is able to detect drifts in gait speed over extended periods of time without any spatial context.

\section{Methodology} \label{sec:methodology}
\subsection{Dataset} \label{sec:dataset}
To rigorously evaluate our floor plan-agnostic approach, we require a dataset with a gait speed drift, which is challenging to obtain in a real-world deployment.
Hence, we adopt a synthetic dataset using the agent-based behavioural simulator developed by Tanaka \textit{et al.}~\cite{tanaka_sensor_2022, tanaka_sensor_2024}. The simulator generates sensor data by modelling an agent's activity and movement across a virtual smart home, through four main components: a floor plan simulator, a behavioural (activity) simulator, a sensor data simulator, and a  walking trajectory simulator. The agent's movement is governed by the latter simulator that generates smooth paths between activity locations by calculating the shortest distance while accounting for the physical layout, discomfort factors related to proximity to furniture or walls, and constraints on abrupt body turning.

Four distinct studio apartment layouts were generated, as they represent the primary configuration options currently provided in the simulator. The sparse sensor placement within these layouts was guided by a product manager from a telecare manufacturer, reflecting typical cost-effective and minimally intrusive commercial installations. An example of these layouts and the resulting walking trajectories can be seen in Figure~\ref{fig:floorplan}.

Nine gait speeds, from $0.4 \; m/s$ to $1.2 \; m/s$ in steps of $0.1 \; m/s$, were simulated. This range was selected to cover speeds indicative of various health states, from frail individuals in acute care (approx. $0.45 \; m/s$) to healthy older adults ($0.94-1.26 \; m/s$)~\cite{peel_gait_2013}. This also includes the critical clinical lower bound of $0.8 \; m/s$, which is predictive of poor outcomes. For our experiments, we simulated a sudden drift in mobility by transitioning from a baseline walking speed to a slower speed after a predefined number of days (\textit{i.e.}, 100 days). To isolate the problem of gait drift detection, we simplified the simulation by excluding other behavioural anomalies permissible by the simulator (\textit{e.g.}, falls or wandering).

The sensors included in the simulation are the following:
\begin{itemize}
    \item PIR Sensors: are activated when the agent's body (modelled with a $0.5 \; m$ radius) enters the sensor's detection area ($1.0 \; m$ radius). They are inactivated when the agent leaves the area or remains still. The sampling rate is $10 \; Hz$, and there is no simulated delay in sending the ``OFF" event.
    \item Pressure Sensors: are activated when the agent is on the sensor mat, with a sampling rate of 10 Hz.
    \item Door Sensors: send binary information, when the door gets opened or closed, capturing as the agent enters or exits the apartment.
\end{itemize}

\subsection{Method}
In contrast to floor plan-dependent approaches~\cite{nait_aicha_continuous_2018, frenken_motion_2011}, our method\footnote{https://github.com/marinavicini/gait-drift-abc26} operates without \textit{a priori} knowledge of the home's floor plan, sensor locations, or optimal walking paths. A core challenge is to identify sensor activation sequences that are reliable indicators of mobility without knowing their physical trajectory.

A sensor event $e_i$ is a tuple consisting of the sensor identifier that fired the event $s$, a timestamp corresponding to when the event was fired $t$, and a categorical value for its status $\sigma \in \{\text{ON}, \text{OFF}\}$:
\begin{equation}
    e = (s, t, \sigma) 
\end{equation}
The method ingests a raw stream of sensor events, considering both `ON' (activation) and `OFF' (deactivation) signals, without differentiating between them.
We aim to identify transitions from one sensor to another, and their respective duration. To do so, we store the sequence of sensors from which the events are coming from ($s_i$, $s_j$), the time difference $\delta$ between consecutive sensor activations in seconds, and the day $\gamma$ of the activation of the last sensor event of this transition:
\begin{equation}
   \tau = (s_i, s_j, \delta, \gamma) \quad \text{with} \, s_i \ne s_j
\end{equation}

To filter out stationary or localised activity, consecutive events from the same sensor identifier (\textit{e.g.}, Sensor A $\rightarrow$ Sensor A) are excluded.
We collect all transitions $\tau$ in the set $\mathcal{T}$.  

Two critical filtering steps are then applied to refine the dataset:
\begin{enumerate}
    \item Minimum Duration Filter ($T_{min}$): Without a floor plan, we cannot know if sensors have overlapping detection fields. This can result in highly frequent sequences with a duration near zero (\textit{e.g.}, $ \delta< 1$ second) that do not represent true ambulation. We apply a minimum duration threshold $T_{min}$ to exclude these non-informative activations.
    \item Maximum Duration Filter ($T_{max}$): Conversely, overly long sequences (\textit{e.g.}, $\delta> 60$ seconds) will unlikely represent a continuous walking path but rather multiple, interwoven activities. We set a maximum duration threshold $T_{max}$ to filter these out.
\end{enumerate}
$$\mathcal{T}_{\text{filtered}} = \{ \tau  \, | \,  T_{min} \leq \delta \leq T_{max}, \: \tau \in \mathcal{T}\}$$

After filtering, we aggregate all transition durations for a given combination of sensors $(s_i, s_j)$ on a given day $\gamma$ in the set $\Delta_{i,j,\gamma}$. Note that we aggregate both directions $(s_i, s_j)$ and $(s_j, s_i)$, making transitions order invariant:
\begin{equation}
\begin{split}
 \Delta_{i,j,\gamma} = \{ \delta \, | \, & \tau = (s_i, s_j, \delta, \gamma) \in \mathcal{T}_{\text{filtered}} \;  \vee  \\
    & \tau = (s_j, s_i, \delta, \gamma) \in \mathcal{T}_{\text{filtered}} \}
\end{split}
\end{equation}

\begin{figure*}
    \centering
    \includegraphics[trim = 80 160 170 185, width=0.75\linewidth]{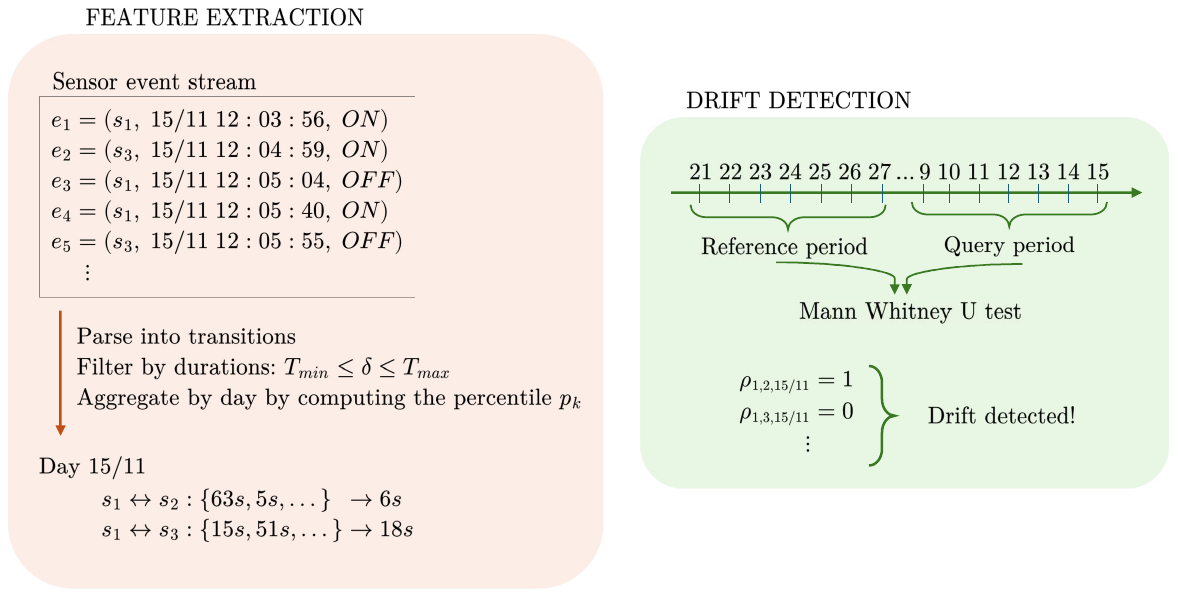}
    \caption{Schematic overview of the proposed floor plan-agnostic gait speed drift detection pipeline. The process consists of two stages: (Left) Feature Extraction, where raw sensor event streams are parsed into sensor-to-sensor transitions, filtered by minimum and maximum durations ($T_{min}, T_{max}$), and aggregated into daily percentile values ($p_k$); and (Right) Drift Detection, where a Mann-Whitney U test compares the distribution of transition durations between a historical Reference Period and a current Query Period to identify statistically significant drifts.}
    \label{fig:example}
\end{figure*}
 
 For each set of transition durations $\Delta$, we now compute a specific percentile $p_k$, that allows for robust estimation, with \textit{e.g.} the 0th percentile (minimum) being a key parameter investigated in our experiments.

\begin{equation}
    p_{i, j, \gamma }^k = \text{percentile}_k \;\Delta_ {i, j, \gamma}
\end{equation}
We count the number of events taken into account to compute the percentile, saving it as the daily support for each sequence $|\Delta_ {i, j, \gamma}|$. A minimum support threshold $S_{min}$ determines if a sequence was traversed a sufficient number of times to be included in the daily calculation.

To detect a drift, we compare a recent \textit{query} period of length $\pi$ (\textit{e.g.}, the most recent week, $\pi =7$) against a stable \textit{reference} period (\textit{e.g.}, the first week of data). For each unique set of sensors, we perform a Mann-Whitney U test~\cite{mann_test_1947}, a non-parametric statistical test, to compare the distribution of its daily durations from the query period against the reference period. 

\begin{align}
P^{query}_{i, j, \gamma} & = \{p^k_{i, j, \gamma - \pi +1}, \, \dots, \, p^k_{i, j, \gamma} \} \\
P^{reference}_{i, j} &= \{p^k_{i, j, 1}, \, \dots, \, p^k_{i, j, \pi} \} \\
U, \text{p-value}_{i,j,\gamma} & =  \text{Mann-Whitney U}(P^{query}_{i, j, \gamma}, P^{reference}_{i, j})
\end{align}

We check if the p-value is below a significance threshold (\textit{e.g.}, $\alpha = 0.05$), indicating a statistically significant drift for that sequence. 
\begin{equation}
\rho_{i, j, \gamma} = \begin{cases} 1 & \text{if } \text{p-value}_{i,j,\gamma} < \alpha \\ 0 & \text{otherwise} \end{cases}
\end{equation}
To aggregate these individual sequence-level tests into a single, robust daily drift decision, we compare two approaches: a simple aggregation and a weighted ensemble. In the weighted ensemble, the drift decisions from all sequences are averaged, weighted by the frequency (support) of each sequence on the last day of the query period, and it is depicted in Equation~\ref{eq:weighted_average}. If a statistical test for a sequence lacks sufficient data (\textit{i.e.}, fails to meet the minimum support $S_{min}$ in the periods), it defaults to a ``No drift" conclusion for that sequence. 

\begin{equation}
\label{eq:weighted_average}
\tilde{\rho}_{\gamma} = \sum_{i < j} w_{i,j, \gamma} *  \rho_{i, j, \gamma} \quad \text{with} \, w_{i, j, \gamma} = \frac{|\Delta_ {i, j, \gamma}|}{\sum_{k < l}|\Delta_ {k, l, \gamma}|} 
\end{equation}
\begin{equation}
\rho_{ \gamma} = \begin{cases} 1 & \text{if } \tilde{\rho}_\gamma \ge 0.5\\ 0 & \text{otherwise} \end{cases}
\end{equation}

\noindent An example of the method can be seen in Figure~\ref{fig:example}.

\begin{figure*}[t]
    \centering
    \begin{subfigure}{0.45\textwidth}
    \centering
    \includegraphics[width=1\textwidth]{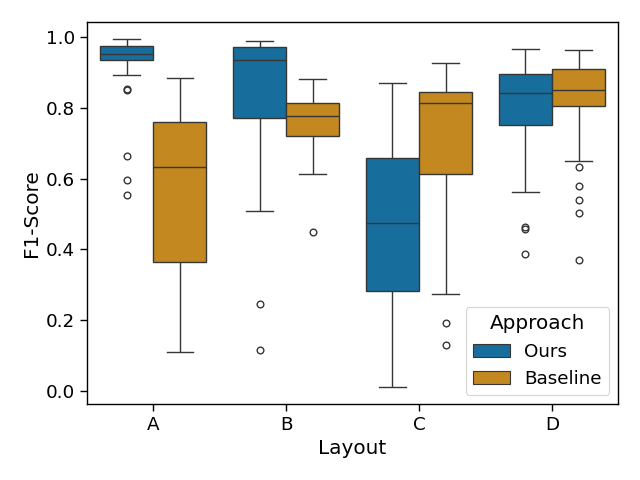}
      \caption{}
      \label{fig:layout_x_f1}
  \end{subfigure}
  \qquad                              
  \begin{subfigure}{0.45\textwidth}
    \centering
    \includegraphics[width=1\textwidth]{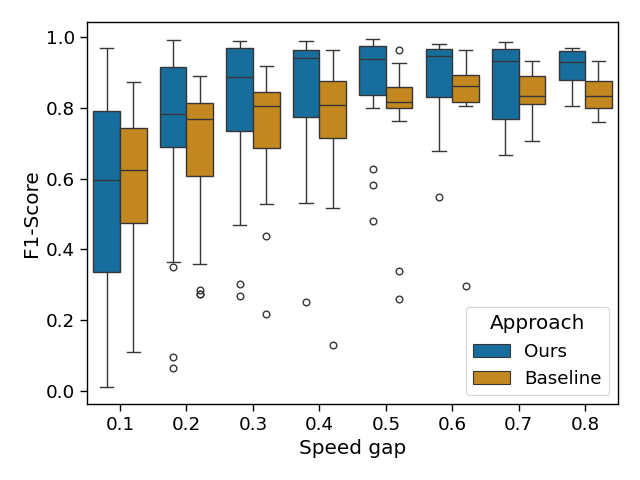}
      \caption{}
      \label{fig:speed_x_f1}
  \end{subfigure}
  \bigskip
  \begin{subfigure}{0.45\textwidth}
    \centering
    \includegraphics[width=1\textwidth]{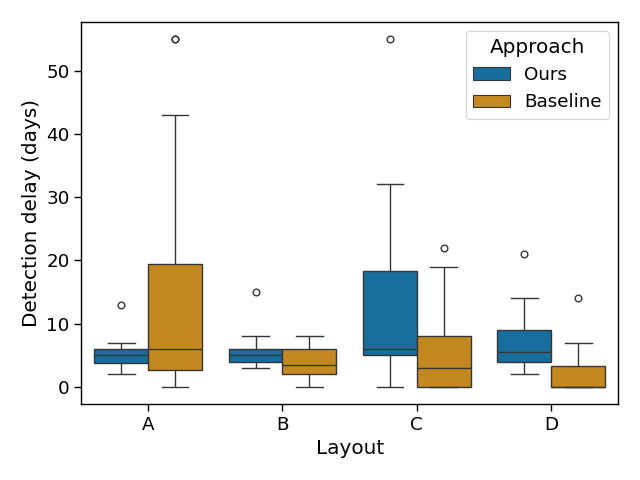}
      \caption{}
      \label{fig:layout_X_delay}
  \end{subfigure}
  \begin{subfigure}{0.45\textwidth}
    \centering
    \includegraphics[width=1\textwidth]{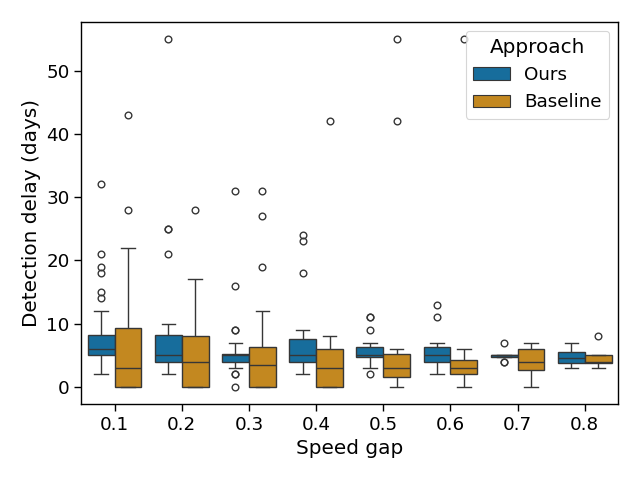}
      \caption{}    
      \label{fig:speedgap_X_delay}
  \end{subfigure}
    \caption{Performance comparison between the proposed floor plan-agnostic method (blue) and the floor plan-dependent baseline (orange) regarding detection accuracy and timeliness. The top row illustrates the F1-Score distribution across (a) the four different apartment layouts (A-D) and (b) varying magnitudes of gait speed drift (Speed gap). The bottom row displays the Detection Delay in days, categorized by (c) the apartment layout and (d) the size of the speed gap.  
    }
    \label{fig:boxplot}
\end{figure*}

\section{Experiments and Results}
We conduct a series of experiments to evaluate our method.

\begin{itemize}
    \item Parameter Sensitivity Analysis: We investigate the effect of varying key hyperparameters: the daily duration percentile ($p_k$), the minimum ($T_{min}$) and maximum ($T_{max}$) duration filters, the minimum support threshold ($S_{min}$), and the impact of applying frequency-based weighting to the ensemble.
    \item Baseline Comparison: We compare the performance (F1-Score and Detection Delay) of our floor plan-agnostic method against the floor plan-dependent baseline~\cite{nait_aicha_continuous_2015, nait_aicha_continuous_2018}. We evaluate this per layout to understand the impact of different home-sensor configurations.
    \item Drift Magnitude Analysis: We assess the method's sensitivity to the magnitude of the gait drift. We hypothesise that detecting small changes (\textit{e.g.}, a $0.1 \; m/s$ drop) is significantly more challenging than detecting large changes (\textit{e.g.}, a $0.8 \ m/s$ drop). We analyse performance across a range of \textit{speed gaps} to determine the method's ability to detect clinically relevant, subtle drifts.
\end{itemize}

\subsection{Baseline}
We benchmark our approach against the work of Nait Aicha \textit{et al.}~\cite{nait_aicha_continuous_2015, nait_aicha_continuous_2018}, a state-of-the-art method that also estimates gait speed from sparse sensors but relies on \textit{a-priori} floor plan information.

Their method requires manually identifying long, straight walking paths from a floor plan. Sensor events are then segmented into trajectories along these predefined paths, with filters applied to isolate movement from stationary activities. To account for variability, they model the observed trajectory durations as a probability distribution (\textit{e.g.}, Poisson mixtures) and use the mean of the best-fit distribution as the representative duration for that path.

Finally, speed is calculated by dividing the path's length, measured from the floor plan, by this estimated mean duration. This dependence on a floor plan to determine path lengths is the primary limitation our work aims to overcome. For a fair comparison, we provide the baseline with these ground-truth path lengths in our experiments.

\subsection{Metrics}
Gait drift detection can be evaluated as a daily binary classification task: for each day, the system must decide whether a significant gait drift has occurred. The following metrics are used:
\begin{itemize}
    \item Accuracy: The proportion of all days that were correctly classified out of the total number of days.
    $$\text{Accuracy} = \frac{TP+TN}{TP + TN + FP + FN}$$
    \item Recall (Sensitivity): The proportion of actual drift days that were correctly identified by the system. 
    $$\text{Recall} = \frac{TP}{TP + FN}$$
    
    \item Precision: The proportion of alerts that were correct out of all alerts issued by the system.
    $$\text{Precision} = \frac{TP}{TP + FP}$$
    \item F1-Score: The harmonic mean of Precision and Recall, capturing the trade-off between the two.
    $$\text{F1-Score} =  2 * \frac{\text{Precision} \times \text{Recall}}{\text{Precision} + \text{Recall}}$$
\end{itemize}

In addition to these standard classification metrics, we introduce a performance indicator for proactive care: 
\begin{itemize}
    \item Detection Delay: This metric measures the timeliness of the detection in days. It is defined as the number of days between the true onset of the gait speed drift and the first day that the system correctly flags an alert. A lower detection delay is highly desirable, as it enables earlier intervention.
\end{itemize}


\subsection{Results}\label{sec:results}
\begin{figure*}[t]
    \centering
    \begin{tabular}{cc}
        \includegraphics[width=0.33\linewidth]{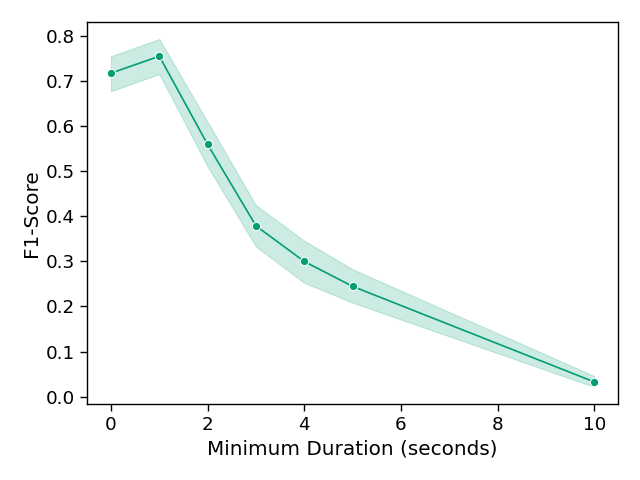} &
        \includegraphics[width=0.33\linewidth]{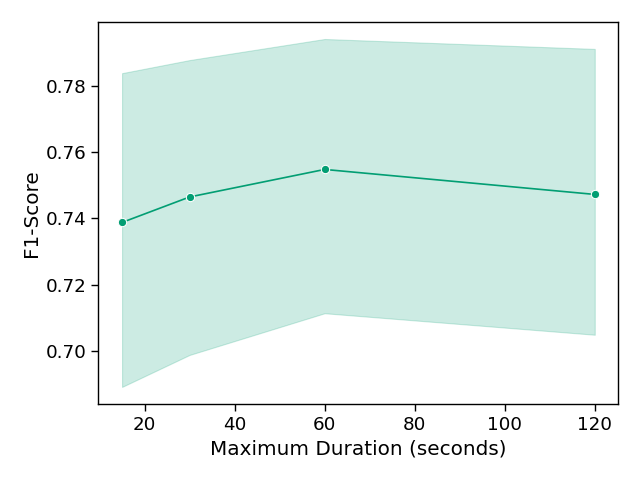} \\
        \includegraphics[width=0.33\linewidth]{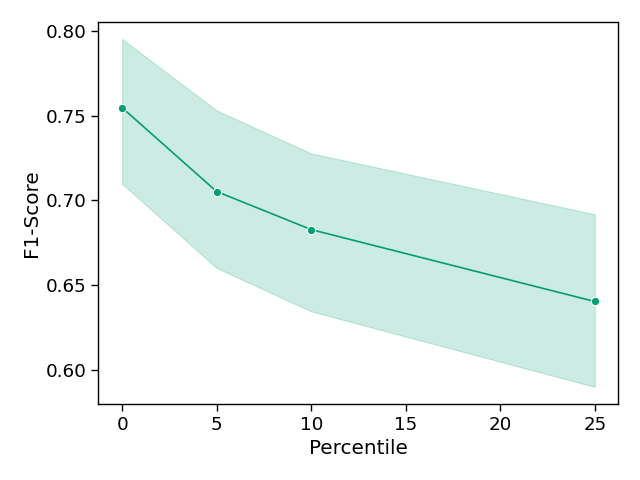} &
        \includegraphics[width=0.33\linewidth]{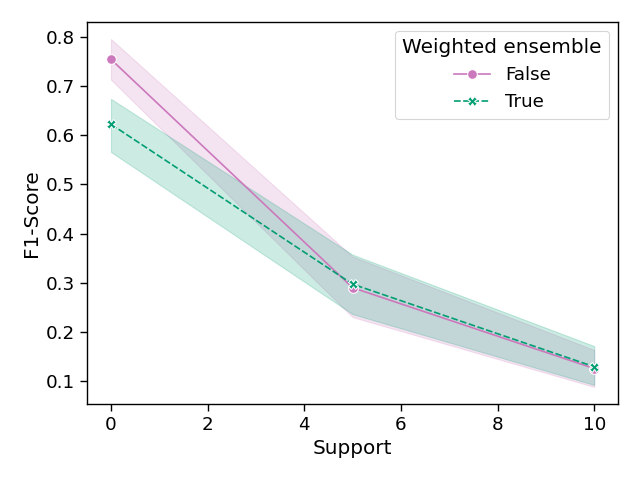}
    \end{tabular}
    \caption{Parameter sensitivity analysis plots. The plots show the impact on F1-Score when varying key hyperparameters. The line indicates the average F1-Score, and the shaded area represents the confidence interval.}
    \label{fig:parameters}
\end{figure*}

The results of the parameter sensitivity analysis provide clear guidance for optimal configuration. Figure \ref{fig:parameters} presents a visualisation of these experiments, illustrating how the F1-Score varies in response to changes in key hyperparameters, specifically the daily duration percentile ($p_k$), minimum/maximum duration filters ($T_{min}, T_{max}$), and minimum support threshold ($S_{min}$).

\subsubsection{Percentile} As shown in Figure~\ref{fig:parameters}, the 0th percentile (\textit{i.e.}, the minimum daily duration) yields the highest F1-Score. This is likely due to the nature of the simulated data, which lacks the noise of real-world environments (\textit{e.g.}, faulty firings, multiple residents, or non-ambulatory movements). In a real-world deployment, a higher percentile (\textit{e.g.}, 10th) might prove more robust by filtering out such noise.

\subsubsection{Duration Filters} Figure~\ref{fig:parameters} illustrates the critical impact of the minimum duration filter. A $T_{min}$ of 0 seconds (no filter) results in poor performance. This is because sequences with near-zero duration, likely caused by overlapping sensor fields (visible in Layout D, Figure~\ref{fig:floorplan}), are included. These sequences are highly frequent but contain no information about mobility, as their duration remains constant regardless of gait speed. Applying a $T_{min}$ of 1 second filters these non-informative sequences, improving performance. As seen in Figure~\ref{fig:parameters}, the maximum duration filter ($T_{max}$) has a negligible impact, suggesting that in our simulation, very long, interwoven sequences do not significantly skew the 0th percentile calculation.

\subsubsection{Support and Weighting} Figure~\ref{fig:parameters} also shows that using a simple, unweighted ensemble provides a slight performance benefit over a frequency-weighted ensemble. Furthermore, we found that the best performance is achieved with a minimum support threshold $S_{min}=0$. Increasing this threshold negatively impacts performance by excluding too many sequences, leading to insufficient data for the statistical tests.

\subsubsection{Baseline Comparison} Fig.~\ref{fig:boxplot} compares our method to the baseline~\cite{nait_aicha_continuous_2018}. Our floor plan-agnostic method achieves a superior F1-Score in layouts A and B, but underperforms in layouts C and D (Figure~\ref{fig:layout_x_f1}). Notably, our method often has a higher detection delay (Figure~\ref{fig:layout_X_delay}), indicating that the baseline's use of optimal, known paths allows it to detect the drift sooner.
The overall performance across the different metrics can be observed in Table~\ref{tab:performance}.

\begin{table}[h]
    \caption{Average performance across all different house layout and different speed gaps combinations. Standard deviation is depicted between parenthesis.}
    \centering
\begin{tabular}{rcc}
              & \textbf{Ours} & \textbf{Baseline~\cite{nait_aicha_continuous_2018}} \\ \hline
\textbf{Accuracy}      & 0.78 \textit{(0.19)}   & 0.69 (0.15)                                    \\
\textbf{F1-score}      & 0.75 \textit{(0.25)}   & 0.72 \textit{(0.20)}                                  \\
\textbf{Precision}     & 0.99 \textit{(0.06)}   & 0.79 \textit{(0.12)}                                    \\
\textbf{Recall}        & 0.66 \textit{(0.28)}   & 0.70 \textit{(0.27)}                                  \\
\end{tabular}
\label{tab:performance}
\end{table}
\subsubsection{Performance vs. Speed Gap} Fig.~\ref{fig:speed_x_f1} analyses performance as a function of the drift magnitude. As hypothesised, the problem is more complex for smaller speed gaps. The F1-Score for both methods degrades as the speed gap decreases, with an average of 0.91 for the larger gap considered (easier problem), and 0.55 for the smallest one (most challenging problem).
Fig.~\ref{fig:speed_x_f1}) shows the baseline method showing an advantage at very small gaps ($0.1-0.2 \ m/s$). This suggests that when the signal is weak, focusing on known, optimal paths (if the floor plan is available) is beneficial, and other sequences can add noise. However, our method remains competitive even at these small gaps, confirming its utility in floor plan-agnostic scenarios.

\section{Discussion and Limitations}

From the results, we see the clear potential for floor plan-agnostic gait speed drift detection in smart homes. This creates significant possibilities for enhancing smart home kits already commercially deployed. Our method's performance was comparable to the baseline, notably outperforming it in Layouts A and C. This suggests that in layouts with less obvious main walking paths, our approach of aggregating all informative sensor-to-sensor transitions is more robust than a method relying on a few predefined optimal paths.

The superior performance of the 0th percentile, as shown in the results, indicates that in a clean simulated environment, the fastest-walked duration is the most stable mobility indicator. We hypothesise this specific parameter may be less robust in real-world deployments, where noise from non-ambulatory movements or multiple residents would likely require a higher, more robust percentile (\textit{e.g.}, 10th) to filter outliers. Moreover, the challenge on detecting smaller speed gaps underlines the inherent difficulty in detecting subtle, though clinically important, changes.

The primary limitation of this work is its evaluation on synthetic data. The simulator, while allowing for a supervised evaluation, does not capture the full complexity of real-world environments. In the data, the agent's walking speed is constant and trajectory variability is low. 
In the paper, we only test with PIR sensors with a 1.0 m radius detection area; we wish to investigate the effects of varied PIR ranges, as our results showed that overlapping sensor fields can create non-informative, near-zero duration sequences that require different filtering logic. 
Furthermore, the simulator does not support multi-residents or visitors, guaranteeing only one person is present. This means our method has not been tested against the significant real-world challenge of data association.

Finally, while the statistical test successfully detects sudden drifts, the reference period is fixed as the initial week. This approach is not designed to detect other types of drifts, such as gradual, recurrent, or incremental decline . Additional research is needed on how to best select the reference period to ensure it does not already contain drifts, perhaps by using a dynamic or rolling window.

\section{Conclusion}
This paper introduces a novel method for detecting gait speed drifts using sparse ambient sensors without any reliance on a floor plan. Our approach successfully aggregates informative sensor transition durations and uses a non-parametric statistical ensemble to detect changes. The method performs comparably to, and in some layouts exceeds, a state-of-the-art baseline that requires full floor plan information, achieving an average F1-Score of 0.75.
This work provides the foundation for scalable and commercially deployable gait monitoring capabilities. Future work must focus on validating this approach in complex, real-world, multi-resident environments and extending the method to detect the more challenging, gradual forms of functional decline.




    


\bibliographystyle{IEEEtran}
\bibliography{Gait_drift}

\end{document}